\documentclass[galaxies,article,accept,moreauthors,pdftex,10pt,a4paper]{mdpi}%
\firstpage{1}
\articlenumber{x}
\doinum{10.3390/------}
\pubvolume{xx}
\pubyear{2016}
\copyrightyear{2016}
\externaleditor{Academic Editors: Jose L. G\'{o}mez, Alan P. Marscher and Svetlana G. Jorstad}
\history{Received: 17 July 2016; Accepted: 12 November 2016; Published: date}

 \theoremstyle{mdpi}
 \newcounter{thm}
 \newcounter{ex}
 \newcounter{re}
 \theoremstyle{mdpidefinition}

\usepackage{soul}

\Title{JVLA Wideband Polarimetry Observations on a~Sample of High Rotation Measure Sources}

\Author{Alice Pasetto $^{1,}$*, Carlos Carrasco-Gonz\'alez $^{2}$, Gabriele Bruni $^{1}$, Aritra Basu $^{1}$, Shane O'Sullivan $^{3}$,  Alex Kraus $^{1}$ and Karl-Heinz Mack $^{4}$}
\AuthorNames{Alice Pasetto, Carlos Carrasco-Gonz\'alez, Gabriele Bruni, Aritra Basu, Shane O'Sullivan,Alex Kraus and Karl-Heinz Mack}

\address{
$^{1}$ \quad Max-Planck Institut f\"ur Radioastronomie (MPIfR),
              Auf dem H\"ugel 69, D-53121 Bonn, Germany; bruni@mpifr-bonn.mpg.de (G.B.); abasu@mpifr-bonn.mpg.de~(A.B.); akraus@mpifr-bonn.mpg.de (A.K.)\\

$^{2}$ \quad Instituto de Radioastronom\'ia y Astrof\'isica (IRyA-UNAM), Antigua Carretera a P\'atzcuaro 8701, 58089~Morelia, Michoac\'an, Mexico; c.carrasco@crya.unam.mx (C.C.-G.)\\
$^{3}$ \quad Instituto de Astronom\'ia (IA-UNAM), Circuito Exterior, \'area de la Investigaci\'on Cient\'ifica, Ciudad Universitaria, Ciudad de M\'exico C.P. 04510 , M\'exico; shane@astro.unam.mx \\
$^{4}$ \quad Istituto di Radioastronomia (IRA-INAF), Via Gobetti 101, I-40129 Bologna, Italy; mack@ira.inaf.it }

\corres{Correspondence: apasetto@mpifr-bonn.mpg.de}

\abstract{We present preliminary results of JVLA wideband full polarization observations of a sample of Active Galactic Nuclei (AGN) with very high Rotation Measure (RM) values, a sign of extreme environment.
Polarization properties show a complex behaviour such that the polarization angle (PA) and fractional polarization (fp) change dramatically within the wide band.
The measured RM is not constant within the wide band. Its complex behaviour reflects the complexity of the medium with the presence of several Faraday components.
The depolarization has been studied by modelling the variations of the Stokes parameters Q and U together with the polarization parameters (PA and fp) with wavelength using combinations of the simplest existing depolarization models.
With this JVLA study we could spectrally resolve multiple polarized components of unresolved AGN.
These preliminary results reveal the complexity of these objects, but improvements to the
depolarization modelling are needed to better understand the polarization structure of these sources.}


\conferencetitle{Blazars through Sharp Multi-Wavelength Eyes. Malaga 2016}

\begin{document}

\section{Introduction}
The polarized non-thermal radiation of the powerful jets of relativistic plasma ejected from radio-loud active galactic nuclei (AGN) can be used to probe the magneto-ionic material along the line of sight between us and the source of emission. The study and analysis of the polarization information in the radio band (i.e., the Faraday rotation and depolarization) are powerful tools that can help to characterize the AGN environment.
The connection of the polarization properties---that is, a very high rotation measure (RM) value and a strong depolarization with the ambient medium---has been studied for years (e.g., \citep[][]{Burn66, Tribble91, Rossetti08}). Both single-dish and
interferometric observations have revealed sources with very high RMs (e.g., \citep[][]{Kravchenko2015, Pasetto2016}),  and these have also been studied with VLBI techniques \mbox{(e.g., \citep[][]{Jorstad07}).}

Here we show JVLA polarimetric measurements of a sample of bright point-like AGN that show one essential characteristic: they are unpolarized at 1.4 GHz in the NRAO VLA Sky Survey (NVSS) \citep{Condon98}. These sources may suffer from strong in-band depolarization; i.e., a large rotation of the polarization angle at this frequency with the result of a final vector pair cancellation of the angles and the subsequent depolarization of the signal. The unpolarized AGN could be polarized at higher frequencies, indicating a very dense medium, possibly also with a strong magnetic field.

We present preliminary results of wideband polarimetric JVLA interferometric observations at~L (1 GHz bandwidth), C, and X bands (each having 4 GHz bandwidth). Thanks to the high-spectral resolution wideband spectropolarimeters, we could follow the dramatic changes of both the fractional polarization and the polarization angle of the targets. We studied the depolarization behaviour by modelling the Stokes parameters Q and U with wavelength (technically the Q/I and the U/I). The~JVLA observations and data reduction are shown in Section \ref{JVLAobs}, a brief description of the depolarization models used in this work is presented in Section \ref{JVLAdepolmodels}, and preliminary results and discussion are presented in Section \ref{JVLAresult}.

\section{Observations and Data Reduction}
\label{JVLAobs}
We observed a sample of 14 sources in full polarization mode by using the Karl. G. Jansky Very Large Array (JVLA) of the National Radio Astronomy Observatory (NRAO; the NRAO is a facility of the National Science Foundation operated under cooperative agreement by Associated Universities,~Inc.).

Observations were made at L (with 1 GHz bandwidth), C, and X bands (with 4 GHz bandwidth). All of the sources are bright enough for phase self-calibration, and are unresolved at all the resolutions of the JVLA. The time on source was set at around 1 minute per source/band; the time was considered to be enough for a good signal-to-noise ratio both for the total intensity and the polarized flux density detection. Observations of a standard flux/polarization angle calibrator (e.g., 3C286, 3C48, 3C138), as well as the leakage calibrators were performed during all the observational sessions.

Data editing and calibration were made by using the data reduction package CASA version 4.4.0 (Common Astronomy Software Applications;  https://science.nrao.edu/facilities/vla/\linebreak data-processing)
following standard VLA procedures (prior known corrections, bandpass, delay, gain calibrations, and finally total flux and polarization calibration).

For the flux calibration of the Stokes I, we used resolved models of the flux calibrators provided by the CASA package. For the calibration of the Stokes parameters Q and U, we used the known values of the fractional polarization and polarization angle at different frequencies reported by \cite{Perley13Pol}. Polynomial functions to these data fit the full Q and U spectrum in the 1 to 45 GHz frequency range. We used those modelling solutions to calibrate our targets (full details on the polynomial functions are reported in \cite{Pasetto2017}).

On the calibrated data, cleaned images of Stokes I, Q, and U were generated for all the targets for each 128 MHz spectral window and for each band.
For the individual spectral window images, we performed Gaussian fits to the brightness distributions, extracting information on the Stokes parameters I, Q, and U.  This high spectral resolution information gives good sampling of the SEDs
~and derived parameters, such as the polarization flux density, the fractional polarization, and the polarization angle for each target.

\section{Depolarization Models}
\label{JVLAdepolmodels}

Total intensity observations of these sources \citep{Pasetto2016} have revealed complex radio spectra, which in most cases could be fitted with multiple synchrotron components. We might therefore also expect a complex behaviour in the polarization information with the presence of multiple interfering RM components. To study the complex polarization behaviour, we fitted the broad band Stokes Q/I and U/I spectra following the procedure proposed by \cite{OSullivan2012}. We used a combination of three simple equations to characterize the depolarization. Depolarization occurs when the medium surrounding the radio source not only produces a change in the polarization angle, but also reduces its polarized flux density (beam
depolarization). The synchrotron emitting and the Faraday rotating regions may also be mixed together (internal depolarization). In order to describe the different depolarization behaviours, we have used three simple known equations: (1) differential Faraday rotation (DFR); (2)~internal Faraday dispersion (IFD); and (3) external Faraday dispersion (EFD) (see \citep[][]{Burn66,Sokoloff98,OSullivan2012}).

All the derivations below start from the following general form of the polarization signal that can be expressed with its complex equation:
\vspace{0pt}
\begin{equation}
p = p_{0}e^{2i(\chi_0+\phi\lambda^2)}
\label{simple}
\end{equation}
with $\phi = K\int^{r}_{0} n_e\vec{B}dl,$ where p$_{0}$ and $\chi_{0}$ are the intrinsic degree of polarization and the intrinsic polarization angle, respectively; $\phi$ is the Faraday depth that depends on the thermal electron density, the magnetic field, and the dimension of the ionized cloud. In the simplest case of Faraday thin objects with a uniform medium and a constant magnetic field, $\phi$ is identical to the RM. The ratio between p and p$_{0}$ represents the depolarization of the polarized signal.
Equation \eqref{simple} represents a single Faraday rotating component.

For a medium which is synchrotron emitting and rotating in the presence of a uniform magnetic field (i.e., the DFR), the complex degree of polarization is given by

\begin{equation}
p = p_0 \frac{\sin \phi \lambda^2}{\phi  \lambda^2} e^{2i\left(\chi_0 + \frac{1}{2} \phi \lambda^2\right)} ,
\label{eqnDFR}
\end{equation}

\noindent where $\phi$ is again the Faraday depth through the region.

If the emitting region also contains a turbulent magnetic field together with a uniform magnetic field (i.e., the IFD), the degree of polarization is then given by

\begin{equation}
p = p_0 e^{2i\chi_0}\left(\frac{1 - e^{-S}}{S} \right) ,
\label{eqnIFD}
\end{equation}

\noindent where $S=2\sigma_{\rm RM}^2 \lambda^4 - 2i\phi\lambda^2$, and $\sigma_{\rm RM}$ is the Faraday dispersion of the random field within the volume traced by the telescope beam.

When the magneto-ionic medium contains a turbulent magnetic field but does not emit synchrotron radiation, the EFD is represented by the equation:

\begin{equation}
p = p_0 e^{-2\sigma_{\rm RM}^2 \lambda^4} e^{2i(\chi_0 + RM \lambda^2)} .
\label{eqnEFD}
\end{equation}

When multiple emitting and/or rotating components exist and they are unresolved within the telescope beam, the complex polarization can be simply described as $p=p_1+p_2 +...+p_N$ \citep{OSullivan2012}. This~is the approach we adopted for the modelling of the sources.
Therefore, we considered only sums of the same model type (i.e., multiple of DFR, multiple of IFD, or multiple of EFD) within the telescope~beam.

The combination of these models has a strong interdependence of the different parameters and contain highly non-linear equations. To evaluate the goodness of the fit, we performed the chi-squared-test to the data using an optimization algorithm for non-linear functions (python's LMfit, which uses the Levenberg--Marquardt method).

\section{Preliminary Results and Discussion}
\label{JVLAresult}
\begin{figure}[H]
	\centering
	\includegraphics[width=0.9\textwidth]{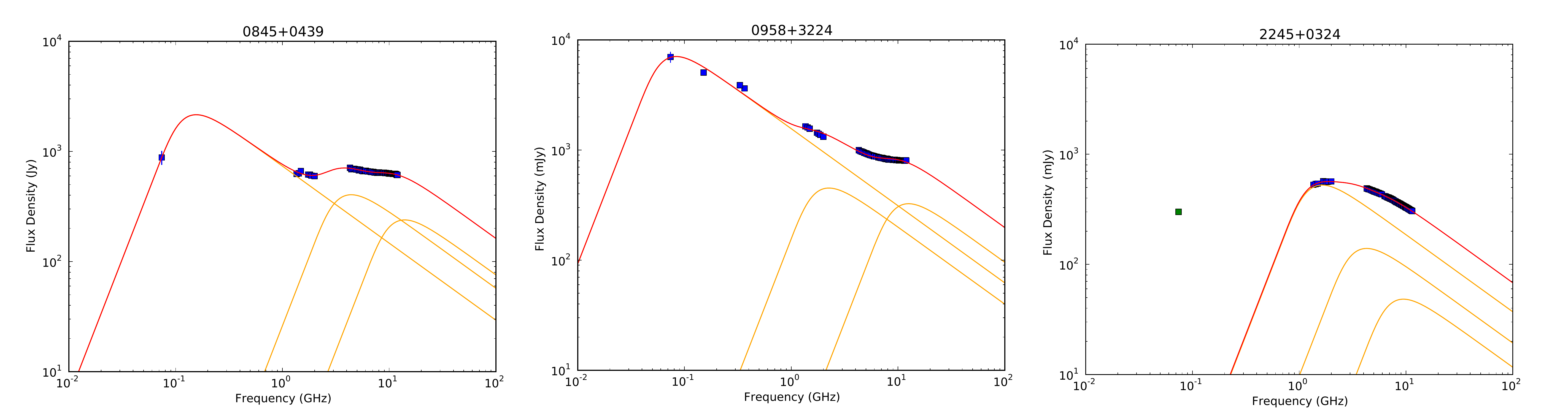}
	\caption{\footnotesize {Radio spectra using L, C, and X bands and literature. Total flux density is expressed in (mJy), and the frequency in (GHz). Blue points are the Karl. G. Jansky Very Large Array (JVLA) and literature data and green points are upper limits. Red line is the sum of the different orange synchrotron components.}}
\label{seds1}
\end{figure}

We present preliminary results of this JVLA observational campaign. A complete description and analysis of these results will be published in \cite{Pasetto2017}.
Examples of the radio spectra for three sources (0845+0439, 0958+3224, and 2245+0324) are shown in Figure~\ref{seds1}. The~radio spectra fits were made by using our JVLA data at L, C, and X bands, as well as data at lower frequencies reported in several surveys (i.e., the VLSS at 74 MHz, the 7C at 151 MHz, the WENSS at 325 MHz and the TEXAS at 365 MHz).
~We fitted the total intensity data with several synchrotron components following a similar approach performed during previous single dish fitting \citep{Pasetto2016} on the same sample. The new results are consistent with previous single dish results.

\newpage
\paperwidth=\pdfpageheight
\paperheight=\pdfpagewidth
\pdfpageheight=\paperheight
\pdfpagewidth=\paperwidth
\newgeometry{layoutwidth=297mm,layoutheight=210 mm, left=2.7cm,right=2.7cm,top=1.8cm,bottom=1.5cm, includehead,includefoot}
\fancyheadoffset[LO,RE]{0cm}
\fancyheadoffset[RO,LE]{0cm}

\unskip

\begin{figure}[H]
	\centering
	\includegraphics[width=0.9\textwidth]{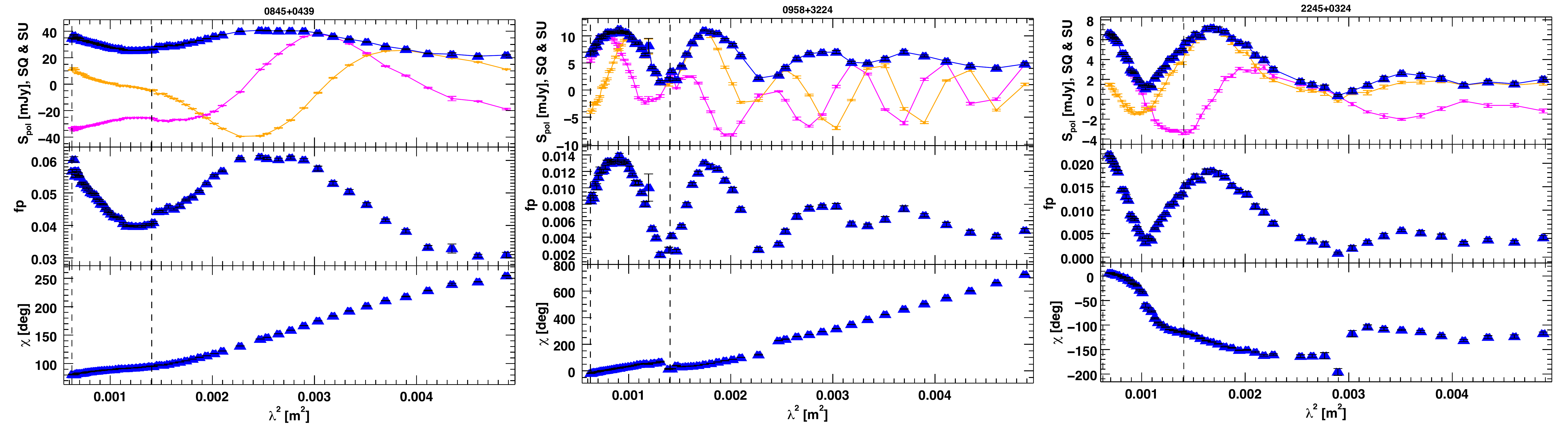}
	\caption{\footnotesize{Polarization measurements of the sources at C and X bands. In the first panel: polarization flux density in blue, orange line is Stokes Q, and pink line is Stokes U; second panel: fractional polarization; third~panel: polarization angle. Vertical dashed lines are the C/X band boundary}}
 \label{pol2}
\end{figure}\unskip


\begin{figure}[H]
	\centering
	\includegraphics[width=0.9\textwidth]{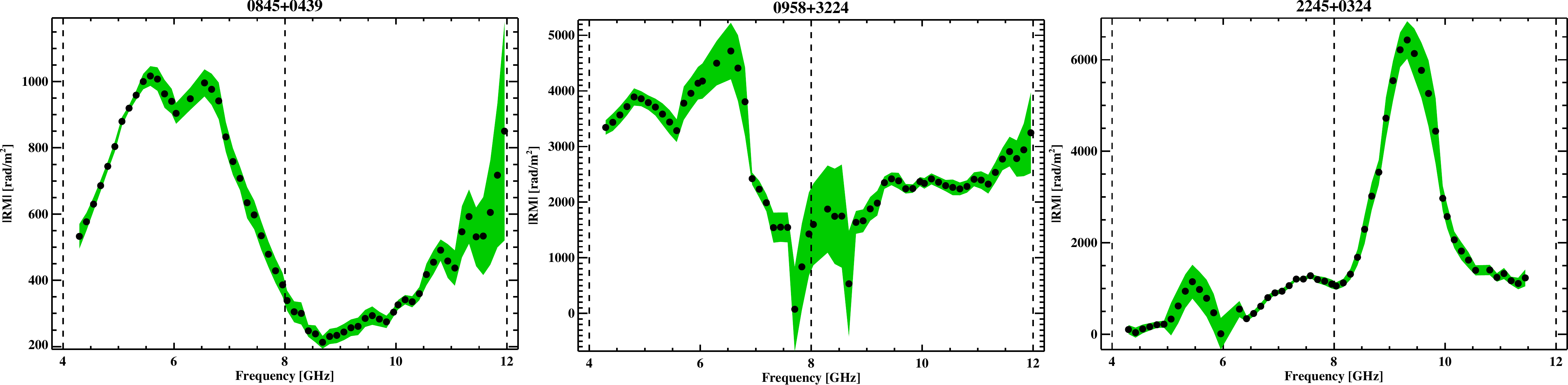}
	\caption{Rotation measure (RM) with frequency behaviour at C and X bands.}
 \label{rmfreq1}
\end{figure}\unskip

\begin{figure}[H]
    \centering
            \includegraphics[width=0.9\textwidth]{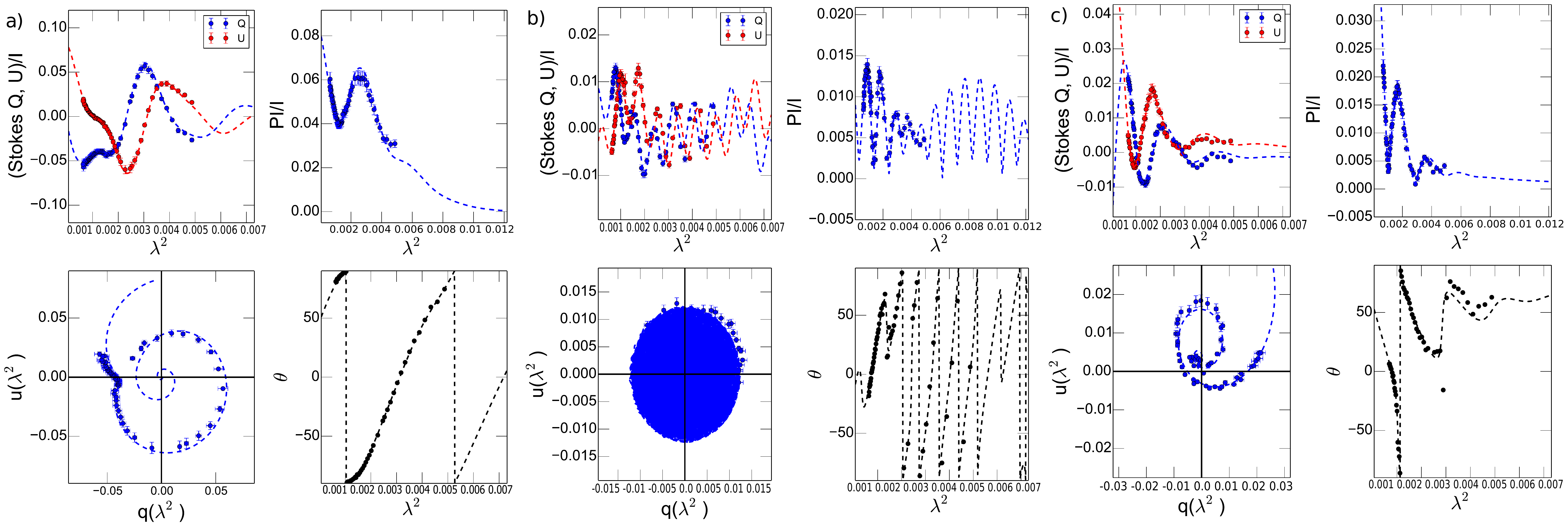}
 \caption{\footnotesize{Depolarization models for the sources: (\textbf{a}) 0845+0439 fitted with a Double External Faraday Dispersion model; (\textbf{b}) 0958+3224 fitted with a Triple model; and (\textbf{c}) 2245+0324 fitted with a Double Internal Faraday Dispersion~model.}}
\label{depol_model}
\end{figure}

\newpage
\restoregeometry
\paperwidth=\pdfpageheight
\paperheight=\pdfpagewidth
\pdfpageheight=\paperheight
\pdfpagewidth=\paperwidth
\headwidth=\textwidth

In Figure~\ref{pol2}, we show the polarization properties of the three sources taken as an example. The~plots show Stokes Q and U together with the polarized flux density \textit{S$_{pol}$}, the fractional polarization fp, and the polarization angle $\chi$ within the C and X bands. These JVLA observations confirm previous results from the Effelsberg campaign \citep{Pasetto2016}: the behaviour of the polarization angle deviates significantly from a simple linear trend (a simple proportionality to $\lambda^2$). We cannot assign a single RM value for these sources in the 4--12 GHz range; at different frequencies, the RM have different values. The variation of the RM with frequency is reported in Figure~\ref{rmfreq1}. We estimated the $d\chi/d\lambda^2$ at each frequency $\nu_0$. The green stripe in the figures represents the 1$\sigma$ error of the derivative (for more details, see \cite{Pasetto2017}).

In Figure~\ref{depol_model}, we report preliminary results of the depolarization modelling for the three sources taken as an example. Our modelling approach could only obtain a reasonable fit for roughly half of the sample. The RM values for the three sources corrected to the rest frame are reported in Table~\ref{RMrestframe}.

\begin{table}[H]
\caption{RM values of the modelling corrected to the rest frame (subscript RF).}
\small
\centering
\begin{tabular}{cccccc}
\toprule

  \textbf{Source}  & \textbf{Model} &   \textbf{z}   &  \boldmath \textbf{RM$_{1RF}$} & \boldmath \textbf{RM$_{2RF}$}&\boldmath \textbf{RM$_{3RF}$}\\
&  &  & \boldmath \textbf{(rad/m$^2$)}  & \boldmath \textbf{(rad/m$^2$)}  & \boldmath \textbf{(rad/m$^2$)}  \\
\midrule

0845+0439    & DEFD   &   0.3 &  1270      $\pm$  10             &  2720      $\pm$ 20       & --         \\
0958+3224    & T	 &   0.5 &  8770      $\pm$  30             &  1670      $\pm$ 110      & 2500 $\pm$ 120   \\
2245+0324	& DIFD   &   1.3 &  --14730    $\pm$  140            &  --21340    $\pm$    590   & --         \\
\bottomrule
\end{tabular}
\begin{tabular}{@{}c@{}}
\multicolumn{1}{p{\linewidth-2cm}}
{\footnotesize NOTE: T: Triple model (3 the simple Equation \eqref{simple}); DEFD: Double External Faraday Dispersion model (2 the EFD model Equation \eqref{eqnEFD}); DIFD: Double Internal Dispersion model (2 the IFD model Equation \eqref{eqnIFD}).}
\end{tabular}

\label{RMrestframe}
\end{table}

\subsection{Some Preliminary Comments on the Sources}

\begin{itemize}[leftmargin=*,labelsep=6mm]
\item Source 0845+0439 (Figures \ref{seds1}, \ref{pol2}, \ref{rmfreq1} and \ref{depol_model}a) \\\vspace{-12pt}

The radio spectrum of this source (Figure~\ref{seds1}) can be fitted with an extended and probably old component at low frequency and two synchrotron components at higher frequencies \citep{Pasetto2016}. However, a flat spectrum over the entire frequency range could also describe the spectrum. The polarization angle clearly does not follow a linear trend (see Figure~\ref{pol2}), with RM changing between 500 and 1000 rad/m$^2$ at C band (see Figure~\ref{rmfreq1}).

This source is very well fitted by a \textit{Double External Faraday Dispersion} model (DEFD, 2 external Faraday rotating screen with a turbulent magnetic field; Equation \ref{eqnEFD}); only a few fractional polarization data points at long wavelength are not well represented by the fit.
From the preliminary results of the depolarization modelling, the magnetic field could be characterized by a high presence of random cells of magnetic field (represented by the $\sigma_{RM}$ values).

A possible scenario could be that the radio synchrotron emission is passing through two turbulent magnetized clouds that are not emitting synchrotron radiation, but that are just responsible for the external Faraday depolarization. These media could be characterized by a high value of thermal electron density.
Depending on the morphology of the source, these clouds on the pc-scale could be attributed to clumps surrounding the central engine and/or a dense wind that covers the inner radio jet of the AGN.
\\
\item Source 0958+3224 (Figures \ref{seds1}, \ref{pol2}, \ref{rmfreq1}, and \ref{depol_model}b) \\\vspace{-12pt}

The spectrum of this source (Figure~\ref{seds1}) could be fitted with an an extended and probably old component at low frequency and two synchrotron components at higher frequencies \citep{Pasetto2016}. The~RM seems to vary within the C and X bands with values $>$ 1000 rad/m$^2$ (see Figure~\ref{rmfreq1}).

For this source, we found a \textit{Triple} model (3 the simple Equation \ref{simple}) to be the best model to fit the data. The depolarization comes from the presence of three external Faraday layers with regular magnetic field. One can visualize it with the presence of at least three clumpy regions, each producing a different RM. The values of the resulting RMs are very high (indeed, note in Figure~\ref{depol_model}b the~large rotation of the Stokes parameters Q and U; see Table \ref{RMrestframe} for the values of the~RM), suggesting a very dense magnetized medium.

This suggests that the radio emission at high frequency comes from the central region of the galaxy, and it goes through at least three external Faraday screens that depolarize at C and X~bands.
\\
\item Source 2245+0324 (Figures \ref{seds1}, \ref{pol2}, \ref{rmfreq1} and \ref{depol_model}c)\\\vspace{-12pt}

This source could be fitted with three synchrotron components \citep{Pasetto2016}, and it is also consistent with a convex shape spectrum (Figure~\ref{seds1}), indicating a possible GHz Peaked Spectrum (GPS) source (considered compact, young radio source, \citep[][]{ODea98}). The RM value increases towards high frequencies, reaching a maximum at $\sim$ 9 GHz, and it decreases again down to an RM of {$\sim$1000~rad/m$^2$} at $\sim$12~GHz (see Figure~\ref{rmfreq1}).

This source is well fitted by a \textit{Double Internal Faraday Dispersion} model (the sum of two IFD model Equation \ref{eqnIFD}): this describes emitting and Faraday-rotating regions with
turbulent and ordered magnetic field components. This model is applied for two Faraday components within the source. If the young nature (i.e., GPS nature) of this source is confirmed (with high angular resolution observations), the internal depolarization would be justified because of the possible presence in this object of both non-thermal (responsible for the synchrotron emission) and thermal (responsible for the Faraday rotation measure) electrons in the same emitting region~\citep{Begelman1999}.
\end{itemize}

With this JVLA study, we could spectrally resolve multiple polarized components of these spatially unresolved AGN.
The preliminary results reveal the complexity of these objects. However, we need to improve the  depolarization modelling to better understand the polarization structure of these sources and possibly give constraints on the physical parameters of the medium; i.e., magnetic field strength and thermal electron density.

\vspace{6pt}
\acknowledgments{This research made use of the NASA/IPAC Extragalactic Database (NED), which is operated by the Jet Propulsion Laboratory, California Institute of Technology, under contract with the National Aeronautics and Space Administration. Alice Pasetto is a member of the International Max Planck Research School (IMPRS) for Astronomy and Astrophysics at the Universities of Bonn and Cologne. {Alice Pasetto and Carlos Carrasco-Gonz\'{a}lez} acknowledge support by UNAM-DGAPA-PAPIIT grant numbers IA101214 and IA102816. }

\authorcontributions{Alice Pasetto conceived and designed the experiments; Alice Pasetto and Carlos Carrasco-Gonz\'{a}lez performed the experiments; Alice Pasetto, Shane O'Sullivan and Aritra Basu analyzed the data; Gabriele Bruni, Alex Kraus and Karl-Heinz Mack contributed reagents/materials/\linebreak analysis tools; Alice Pasetto wrote the paper.}

\conflictofinterests{The authors declare no conflict of interest.}

\bibliographystyle{mdpi}

\renewcommand\bibname{References}

\end{document}